\documentclass[journal=jacsat,manuscript=article]{achemso}

\usepackage{chemformula} 
\usepackage[T1]{fontenc} 

\author{Ravindra Kumar Yadav}
\email{ryadav@ccny.cuny.edu}
\affiliation{Department of Physics, The City College of New York, 85 St. Nicholas Terrace, 85 St. Nicholas Terrace, New York, 10031,  USA.}
\author{Sitakanta Satapathy}
\affiliation{Department of Physics, The City College of New York, 85 St. Nicholas Terrace, 85 St. Nicholas Terrace, New York, 10031,  USA.}
\author{Prathmesh Deshmukh}
\affiliation{Department of Physics, The City College of New York, 85 St. Nicholas
Terrace, 85 St. Nicholas Terrace, New York, 10031,  USA.}
\altaffiliation{The PhD Program in Physics, Graduate Center of the City University of New York, 365 5th Ave, New York, 10016, USA.}
\author{Biswajit Datta}

\affiliation{Department of Physics, The City College of New York, 85 St. Nicholas
Terrace, 85 St. Nicholas Terrace, New York, 10031,  USA.}
\author{Addhyaya Sharma}
\affiliation{Department of Physics, The City College of New York, 85 St. Nicholas
Terrace, 85 St. Nicholas Terrace, New York, 10031,  USA.}
\author{Andrew Olsson}
\affiliation{Department of Chemistry, Indiana University,  IN 47405,
 USA }
\author{Junsheng Chen}
\affiliation{Nano-Science Center and Department of Chemistry,  University of
Copenhagen,  Denmark.
}
\author{Bo W. Laursen}
\affiliation{Nano-Science Center and Department of Chemistry,  University of
Copenhagen,  Denmark.
}
 \author{Amar H. Flood}
\affiliation{Department of Chemistry, Indiana University,  IN 47405,
 USA }

\author{Matthew Y. Sfeir}
\affiliation{Photonics Initiative, Advanced Science Research Center, City University of New York, New York, 85 St. Nicholas Terrace, New York, 10031,
 USA.}

\author{Vinod M. Menon}
\email{vmenon@ccny.cuny.edu }
\affiliation{Department of Physics, The City College of New York, 85 St. Nicholas
Terrace, 85 St. Nicholas Terrace, New York, 10031,  USA.}
\altaffiliation{The PhD Program in Physics, Graduate Center of the City University
of New York, 365 5th Ave, New York, 10016, USA.}

\title[An \textsf{achemso} demo]
  {Direct writing of room temperature polariton condensate lattice by top-down approach} 


\begin{document}


\begin{abstract}
 Realizing lattices of exciton polariton condensates has been of much interest owing to the potential of such systems to realize analog Hamiltonian simulators and physical computing architectures. Prior work on polariton condensate lattices has primarily been on GaAs-based systems, with the recent advent of organic molecules and perovskite systems allowing room-temperature operation. However, in most of these room temperature systems, the lattices are defined using a bottom-up approach by patterning the bottom mirrors, significantly limiting the types of lattices and refractive index contrast that can be realized. Here, we report a direct write approach that uses a Focused Ion Beam (FIB) to etch 2D lattice into a planar microcavity. Such etching of the cavity allows for realizing high refractive index contrast lattices. We realize the polariton condensate lattice using the highly photostable host-guest Frenkel excitons of an organic dye small molecular ionic lattice  (SMILES)\cite{benson2020plug,deshmukh2023plug}. The lattice structures are defined on a planar microcavity embedded with SMILES using FIB, allowing the realization of lattices with different geometries, including defect sites on demand. The band structure of the lattice and the emergence of condensation are imaged using momentum-resolved spectroscopy. The present approach allows us to study periodic, quasi-periodic, and disordered polariton condensate lattices at room temperature using a top-down approach without compromising on the quantum yield of the organic excitonic material embedded in the cavity.

\end{abstract}

\section{Introduction}
Exciton-polaritons are hybrid quasiparticles that arise from the strong coupling between excitons in semiconductors and cavity photons in microcavities. These polaritons possess a small effective mass inherited from the photonic component, as well as strong interactions stemming from the excitonic part\cite{byrnes2014exciton,keeling2020bose,pandya2021microcavity,dufferwiel2015exciton}. The combination of small effective mass, interaction effects, and bosonic nature enables the observation of intriguing phenomena like Bose-Einstein-like condensation, superfluidity, and vortex formation\cite{byrnes2014exciton,altman2015two,amo2009superfluidity,lagoudakis2017polariton,maitre2020dark,lagoudakis2008quantized}. The realization of exciton-polariton condensate lattices has emerged as an attractive platform for analog Hamiltonian simulators, such as the classical XY Hamiltonian, and for physical computing architectures\cite{lagoudakis2017polariton,berloff2017realizing,amo2016exciton,kavokin2022polariton,barik2018topological}.
Previous studies on polariton condensate lattices have primarily focused on GaAs-based systems\cite{jacqmin2014direct,galbiat,lai2007coherent,whittaker2018exciton,alyatkin2020optical}, with recent advancements in organic molecules and perovskite systems allowing for room temperature operation \cite{yagafarov2020mechanisms,daskalakis2015spatial,betzold2019coherence,plumhof2014room,su2020observation,su2018room,su2021optical,su2021perovskite,wu2021nonlinear,spencer2021,tao2022halide}. However, most of these room temperature systems rely on a bottom-up approach, where the lattices are defined by patterning the bottom mirrors\cite{dusel2020,scafirimuto2021tunable,jayaprakash2020two,su2020observation}. This is due to the detrimental effect of top-down approaches on the condensates and, thereby significantly limiting the types of lattices and the index contrast that can be achieved.
Here, we present a top down approach that utilizes Focused Ion Beam (FIB) etching to pattern a planar microcavity. This etching technique enables the realization of high-contrast lattices, as previously demonstrated in GaAs systems\cite{jacqmin2014direct,tao2022halide,real2020semi}.
Specifically, we report direct writing of a two-dimensional exciton-polariton condensate lattice by patterning an array of overlapping pillars etched into a planar microcavity. The patterned microcavity structure shows condensation at room temperature along with the bands arising from the lattice structure. To characterize the fabricated lattice, we employ energy-resolved wavevector imaging of photoluminescence (PL) and perform power-dependent PL measurements to study optical nonlinearities. 
\section{Results and discussion}

\begin{figure}[h]%
\centering
\includegraphics[width=0.9\textwidth]{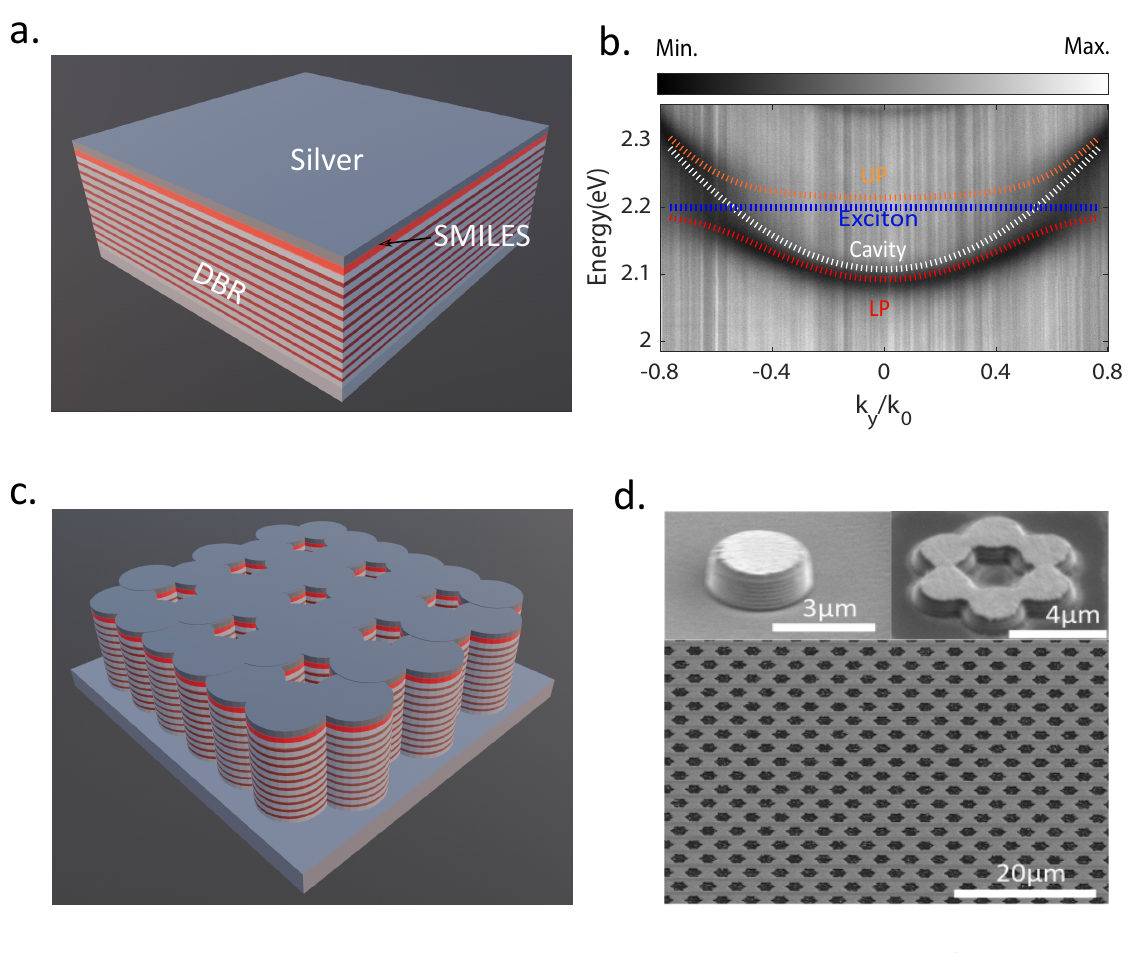}
\caption{(a) Schematic  for planar organic exciton-polariton cavity. (b) White light angle-dependent reflectivity of the planar cavity region along with coupled oscillator model fits to the observed dispersion. (c) Schematic  for patterned honeycomb organic exciton-polariton lattice . (d) Scanning electron microscope image of a single micropillar,  honeycomb unit cell, and lattice patterned on the planar cavity using FIB.}
\label{Fig.1}
\end{figure}
Here, we fabricate an exciton-polariton microcavity on a glass substrate, as shown in Fig.~\ref{Fig.1} a. The cavity consists of a 10.5-pair bottom distributed Bragg reflector (DBR) consisting of alternating layers of a quarter wavelength thick $SiO_2$ and $TiO_2$ deposited via radio-frequency sputtering. The center wavelength of the bottom DBR was designed to be around 600 nm, and the bandwidth was 200nm. We use the recently introduced an organic dye small molecule ionic isolation lattice (SMILES) system as the excitonic material owing to their enhanced photostability and high quantum yield \cite{benson2020plug,kacenauskaite2022universal,deshmukh2023plug}. Specifically, we use rhodamine 3B (R3B), a widely used laser dye (with bandgap  2.19 eV,  566 nm), composed as a SMILES material (See methods)(Supporting Information(SI), Fig.S1). $30nm$ R3B-SMILES in PMMA matrix is deposited via spin coating onto the bottom DBR as shown in Fig.\ref{Fig.1}a. The cavity is completed by depositing $~100nm$ thick silver film on top of the SMILES layer. This results in the formation of a Tamm plasmon cavity mode with loaded Q of $200$ similar to our previous report\cite{deshmukh2023plug}. Strong coupling between the Tamm mode and the excitons of the SMILES results in the formation of polariton states whose dispersion obtained via Fourier imaging is shown in Fig.\ref{Fig.1}b along with fits based on the coupled oscillator model. Anticrossing between the upper(UP) and lower polariton(LP) branches is observed at finite in-plane momentum due to the negative detuning ($40meV$) of the cavity with respect to the exciton in SMILES. The estimated Rabi splitting based on the coupled oscillator model fit is $100meV$.
The strongly coupled cavity is patterned using FIB etching with low ion current ($77pA$) and $30kV$ voltage to avoid heating caused by the gallium ion beam. 
The microcavity is patterned into different structures, such as a single pillar, honeycomb unit cell, and honeycomb lattice, as shown schematically in  Fig. \ref{Fig.1}c. The scanning electron microscope images of the patterned structures are shown in  Fig. \ref{Fig.1}d.

In the single pillar case, the refractive index contrast arising from the etching of the pillar (diameter= $2.75\mu$m) results in lateral confinement and emergence of discrete polariton states\cite{ferrier2011,mangussi2020mult,klaas2018,kuznetsov2018,klein2015polariton,galbiati} as opposed to dispersive branches observed in planar cavities as shown in SI, Fig.S2a.  
When the excitation power is increased above the condensation threshold ($1.25p_{th}$) the emission is observed from the lowest energy state as seen in SI, Fig.S2b. 
\begin{figure}[H]%
\centering
\includegraphics[width=0.9\textwidth]{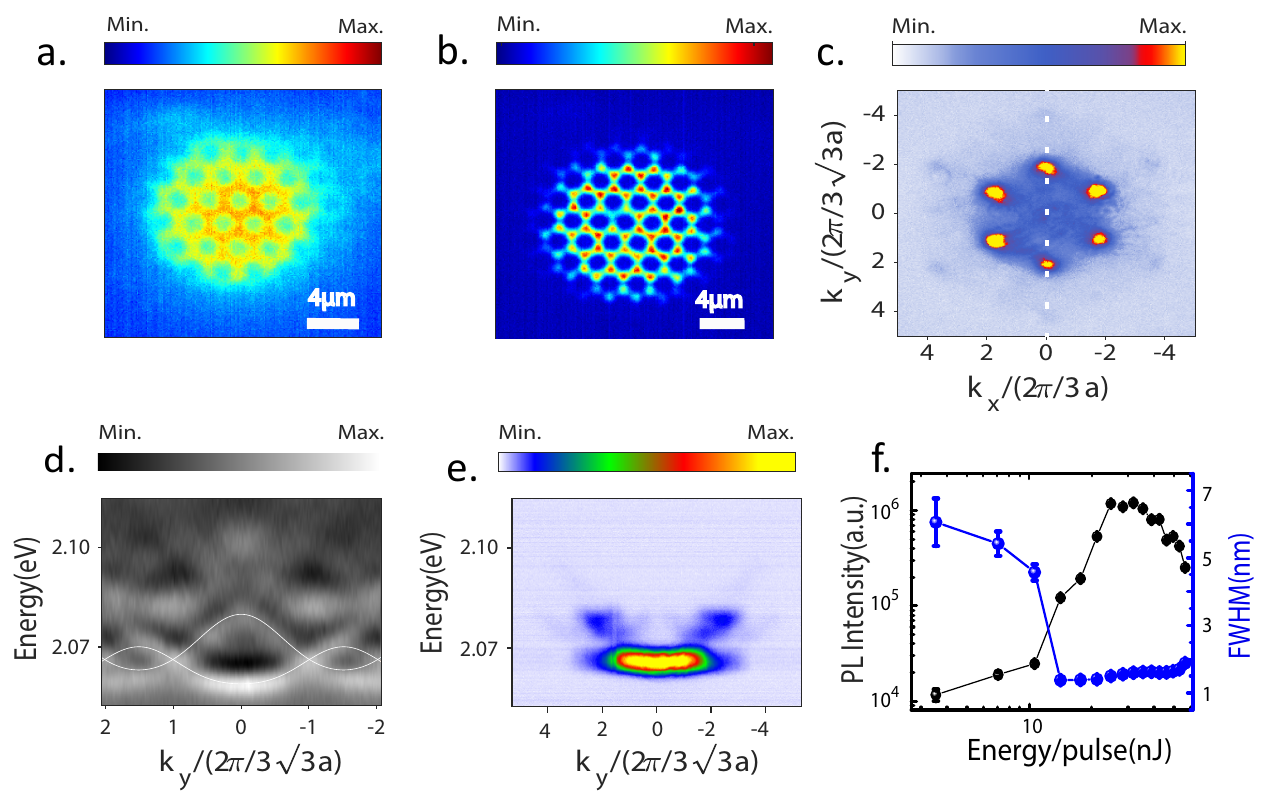}
\caption{  Measured real space PL image of honeycomb lattice  (a) below condensation threshold and (b) above condensation threshold. (c) Measured momentum space PL emission image of condensate lattice above condensation threshold showing the condensation at the corners of the BZ. Energy resolved momentum space  PL  (d) below condensation threshold along with tight binding model fits to the observed dispersion and (e) above condensation threshold.(f)Integrated PL intensity and full width at half maximum (FWHM) of condensate lattice PL as a function of pulse energy.}\label{fig3}
\end{figure}

Two-dimensional array of micropillars in a honeycomb lattice geometry with micropillar diameter(\textit{d}) $2.75\mu$m and lattice constant(\textit{a}) $2.40\mu m$ is realized using FIB following similar approach to the single pillar. The polariton lattice is excited nonresonantly at $514$ nm wavelength using a flat top beam with a spot size of approximately 25$\mu m$, generated by tightly focusing 100 femtosecond pulsed laser with a repetition rate of $1KHz$ on the back focal plane of a 50x microscope objective (numerical aperture=0.80). Fig.\ref{fig3}a and \ref{fig3}b illustrate the real-space PL of the lattice below and above the condensation threshold (pulse energy=10nJ), respectively. A distinct transition is observed from delocalized PL to localized PL, specifically at the lattice sites. 
Fig.\ref{fig3}c displays the momentum resolved image of the PL from the lattice after condensation, demonstrating the localization of polariton PL at the six Dirac points of the first Brillouin Zone(BZ). Fig.\ref{fig3}d shows the experimentally observed band structure of the honeycomb lattice in Energy resolved momentum space  PL  with the momentum slice taken along $k_x$=0 (dashed line in Fig.\ref{fig3}c) below the condensation threshold. Modified bandstructure along other in-plane momentum directions are shown in SI Fig.S3a-c.Above the threshold, we observe the collapse of PL  to the lowest energy bands as shown in Fig.\ref{fig3}e. Pump power dependence of emission intensity and linewidth is shown in Fig.\ref{fig3}f. Clear threshold and nonlinear increase in output intensity is observed along with the decrease in the linewidth indicative of the onset of polariton condensation similar to what was reported previously  in non-patterned SMILES system\cite{deshmukh2023plug}.

\begin{figure}[h]%
\centering
\includegraphics[width=1.00\textwidth]{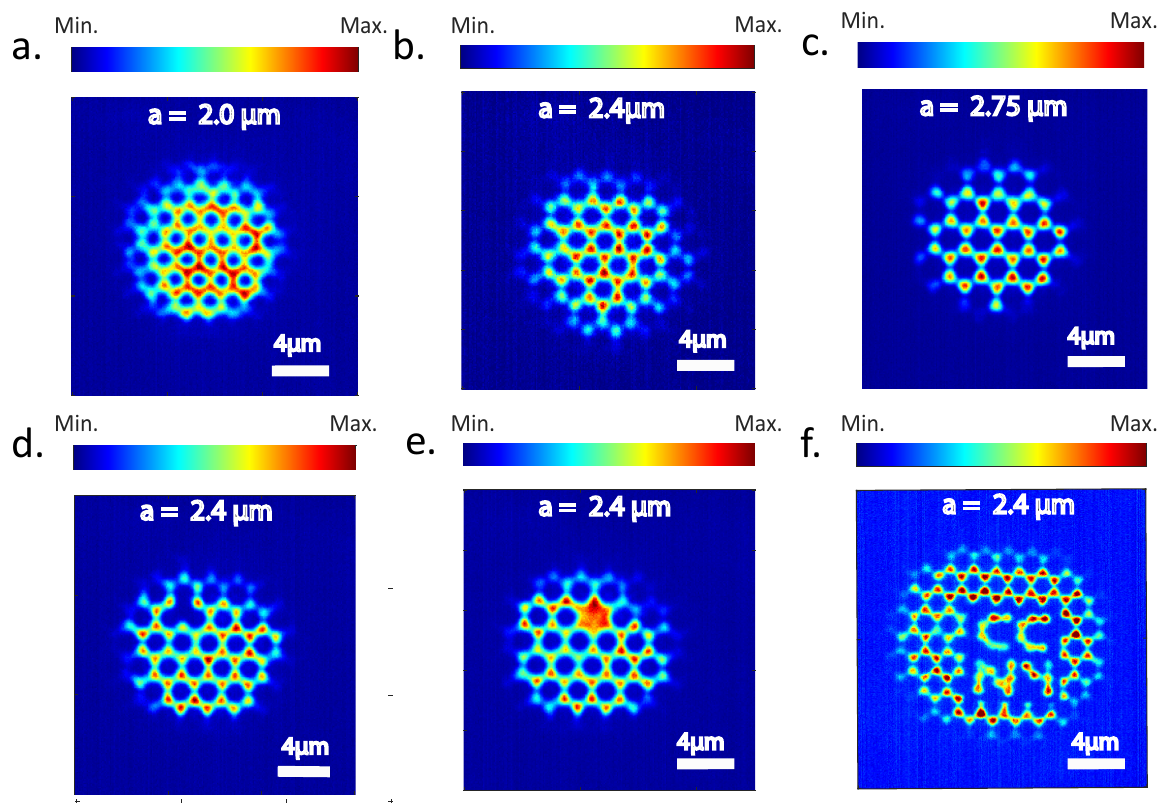}
\caption{ Real space PL  image of polariton condensate lattice with micropillar diameter \textit{d}=$2.75\mu m$, and lattice constant  (a) $a=2.0\mu m$, (b) $a=2.4\mu m$, (c) $a=2.75\mu m$,(d) $a=2.4\mu m$ with missing lattice sites defect,(e) $a=2.4\mu m$ with missing hole,(f) $a=2.4\mu m$ with "CCNY" logo as an example of arbitrary defects that can be realized through design. }\label{fig4}
\end{figure}

\begin{figure}[H]%
\centering
\includegraphics[width=0.9\textwidth]{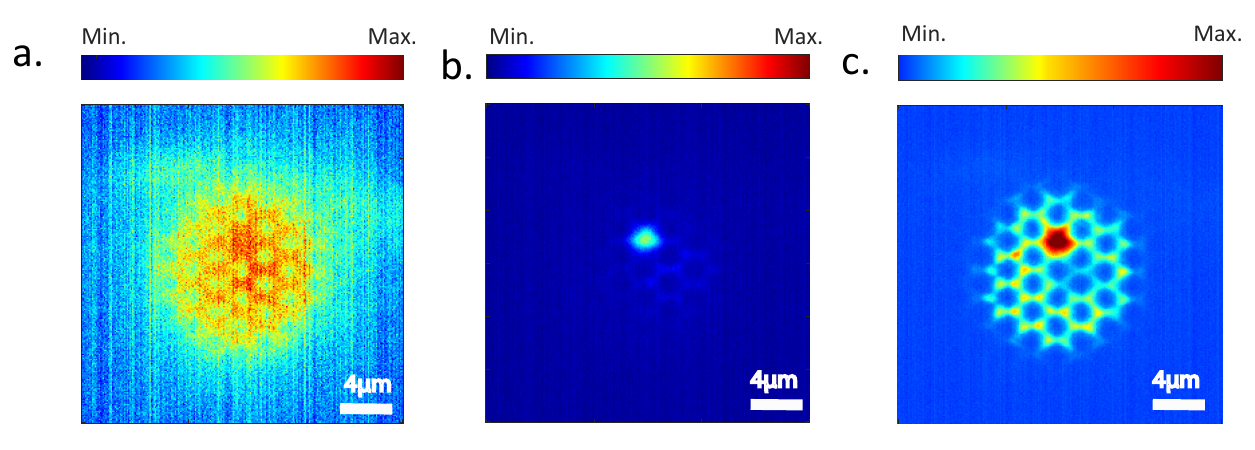}
\caption{  Real space PL image of defect lattice with missing hole at  (a) $p=0.5 p_{th}$ collected for longer integration time ($30sec$) to show background lattice below condensation threshold, (b) $ p=p_{th}$ collected with integration time $1sec$ and (c) $ p=1.25p_{th}$ collected with integration time $1sec$. Preferential condensation at the lattice defect site is observed.}\label{fig5}.\end{figure}

Finally, we investigate polariton condensation in 2D honeycomb lattices with varying degree of coupling between the lattice sites and on the presence of defects.  Fig.\ref{fig4}a-c, show the real space image of condensation in lattices with varying degree of coupling between lattice sites with lattice constant varying from  $a=2.0\mu m$ to $2.75\mu m$. We observe the transition from delocalized to localized condensate formation by increasing the inter-pillar distance (reduced intersite coupling). For the highly coupled case (Fig.\ref{fig4}a), the condensate is delocalized over the entire structure in contrast to the weakly coupled case (Fig.\ref{fig4}c), where the condensates are localized at the pillar sites. The versatility of our top-down approach is shown in Fig.\ref{fig4} d-f, where lattices with a missing pillar (Fig.\ref{fig4}d), missing hole (Fig.\ref{fig4}e) and arbitrary defect pattern showing the abbreviation of City College of New York (CCNY) (Fig.\ref{fig4}f) are realized. We also studied the role of lattice defects in the condensate formation. Shown in Fig.\ref{fig5} a-c is the real space image of the PL from the lattice for increasing pump power. Below the condensation threshold, the PL is distributed throughout the lattice. The onset of condensation is observed first at the defect site (Fig.\ref{fig5}b), followed eventually by condensation emerging across the entire lattice. The threshold for condensation at the lattice site is lower than what was observed previously in defect-free lattices, indicating enhanced scattering to the defect site. 

 \section{Conclusion}
 In summary, we report the realization of room temperature polariton condensate lattice using a top-down fabrication approach involving FIB etching on the host-guest-based organic SMILES material microcavity. Experimental results demonstrate the confinement of polaritons resulting in discrete states and condensation to the ground state energy in the energy-and momentum resolved spectroscopy of single micropillars. Through the patterning of a honeycomb lattice on the SMILES microcavity, we have successfully realized a polariton condensate lattice consisting of coupled micropillars. By changing the lattice constant, we demonstrate the transition from delocalized to localized condensate. Furthermore, we study the effect of deterministically placed defects in the condensate lattice, revealing that condensates prefer high refractive index sites for localization and also reduce the condensation threshold pump power.
Our top-down approach provides a versatile platform for studying arbitrary polariton condensate lattices and polaritonic circuits at room temperature over large area.
\section{Methods}\label{sec11}
\subsection{SMILES Microcavity}To prepare the SMILES solution, we followed a previously reported recipe\cite{benson2020plug,deshmukh2023plug}. The SMILES-based microcavity is fabricated on a quartz substrate with a $500 nm$ thick distributed Bragg reflector (DBR) centered at $620nm$. The DBR, composed of 10.5 pairs of $SiO_2$/$TiO_2$, was cleaned using $O_2$ plasma for $5$ minutes. A $30nm$ thick SMILES film is then deposited on the cleaned DBR using a spin coating technique with a two-step process. In the first step, the film is spun for $20$ seconds at a speed of $1000$ rpm, followed by a second step of spinning for $80$ seconds at $3000$ rpm. The deposited SMILES film is placed in a constant pressure vacuum at $25$ \textcelsius. Finally, a $100$ nm thick silver layer is deposited on top of the SMILES film using an e-beam evaporator, completing the fabrication of the microcavity.

\subsection{Experimental techniques}We employed real space PL and Fourier space PL imaging to map the condensate emission characteristics and the band structure of the lattice. For our measurements, we utilized two lasers. To investigate the band structures, we used a 488 nm laser with a repetition rate of 76 MHz to excite the patterned structure on the silver sides of the microcavity. For the excitation of the larger area in the lattice, we created a large laser spot ($~25\mu m$) by focusing the laser on the back focal plane of the objective lens. The band structure of the lattice is measured using far-field imaging techniques, where the back focal plane of the objective lens is imaged in front of a charged coupled device(CCD) camera. Energy-resolved angle-dependent PL was measured by selecting a specific in-plane momentum using a narrow slit in front of the CCD camera and dispersing the PL using a $300$ grooves per line grating. In addition, we employed a pulsed $514$nm laser with a pulse width of $280$ fs and a repetition rate of $1 kHz$ for the condensation experiments. The Brillouin zone and band structure were measured using the Fourier imaging technique.
\subsection{Coupled oscillator model} Upper and lower polariton branches in reflectivity of the strongly coupled cavity  are fit using coupled oscillator model defined by following Hamiltonian matrix:
\\
\\
\begin{equation}
  H=
\begin{bmatrix}
 E_x(k)-\iota \gamma_x/2&g&\\
g&E_c(k)-\iota \gamma_c/2&\\   
\end{bmatrix}  
\end{equation}
\\
\\
Where $g$ is the coupling strength between exciton and single cavity mode. $E_x(k)$ and $E_c(k)$ are the energy of exciton and cavity mode, respectively.$\gamma_x$ and $\gamma_c$ are full width at half maximum(FWHM) of exciton and cavity mode resonance, respectively.
\subsection{Tight binding model for bandstructure} Tight binding approximation is used to fit the bandstructure of the honeycomb polariton lattice. The tight binding model for honeycomb polariton lattice is given by the following expression\cite{jacqmin2014direct,pan2019two,mangussi2020mult}:
\begin{equation}
    E(k)=\pm t_{n}\sqrt{3+F(k)}-t_{nn}F(k)
\end{equation}
where F(k) is expressed as:
\begin{equation}
    F(k)=2\cos{\sqrt{3}k_ya}+4\cos{\frac{\sqrt{3}}{2}k_ya}.\cos{\frac{3}{2}k_xa}
\end{equation}
where $t_n$ and $t_{nn}$ are coupling to the first and second nearest neighbor and used as free parameters in the fitting. Extracted value of $t_{nn}$/ $t_{n}$ is $\approx 0.10$, which is in good agreement with the previously reported pattern structure through etching\cite{jacqmin2014direct}.
\section{\textcolor{black}{Supporting Information}}
 \textcolor{black}{Absorption and photoluminescence of SMILES, Polariton condensation in single micropillar, Bandstructure of honeycomb polariton lattice.} 
\section{Acknowledgments}V.M.M. and R.K.Y. were supported by the U.S. Air Force Office of Scientific Research$-$MURI Grant FA9550$-$22$-$1$-$0317. S.S., and P.D., acknowledge support from the US National Science Foundation (NSF$-$ QTAQS program OMA$-$1936351. M.Y.S. work was supported by the U.S. Department of Energy, Office of Science, Office of Basic Energy Sciences under Award No. DE$-$SC0022036. A.H.O. and A.H.F. acknowledge support from the U.S. National Science Foundation (DMR-2118423).
 \section*{Declarations}Financial Declaration: The authors declare the following competing financial interest(s): A.H.F. and B.W.L. are cofounders in Halophore, Inc. R.K.Y., S.S., P.D., V.M.M., B.W.L., and A.H.F. have filed a provisional patent application.
\section*{Data Availability}
All data will be provided by the corresponding authors upon reasonable request.

\providecommand{\latin}[1]{#1}
\makeatletter
\providecommand{\doi}
  {\begingroup\let\do\@makeother\dospecials
  \catcode`\{=1 \catcode`\}=2 \doi@aux}
\providecommand{\doi@aux}[1]{\endgroup\texttt{#1}}
\makeatother
\providecommand*\mcitethebibliography{\thebibliography}
\csname @ifundefined\endcsname{endmcitethebibliography}
  {\let\endmcitethebibliography\endthebibliography}{}

\appendix
\clearpage
\centerline{Supporting Information}
\setcounter{figure}{0}
\section{Absorption and photoluminescence of SMILES }\label{sec1}
\begin{figure}[h]%
\centering
\includegraphics[width=0.9\textwidth]{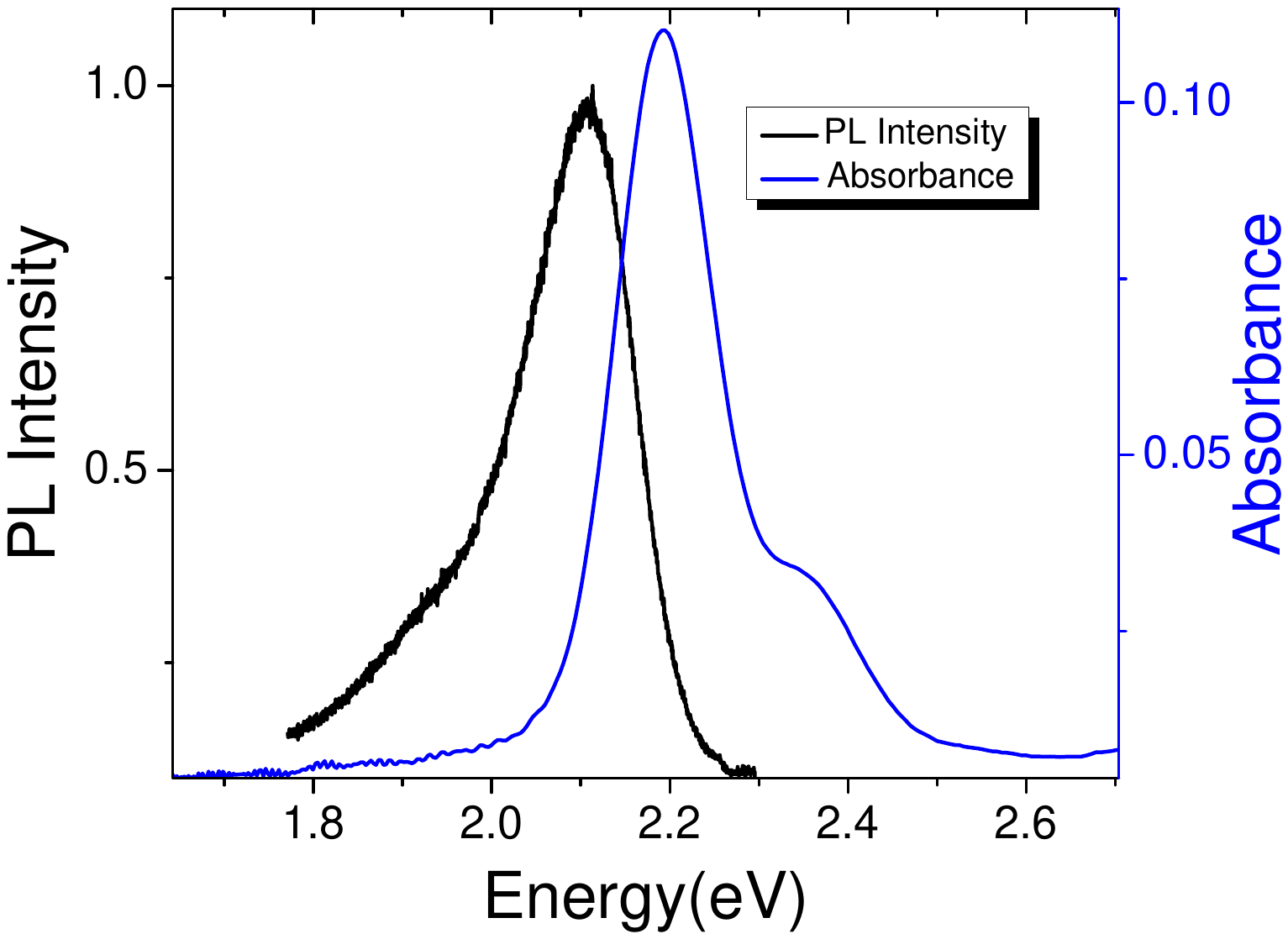}
\caption{ Absorbance and normalized PL intensity of SMILES film on glass}\label{figS2}
\end{figure}

Fig.\ref{figS2} shows absorption and photoluminescence of thin SMILES film on glass. Absorption spectra are measured using a UV-Vis spectrometer, and the PL of the thin film was collected by exciting SMILEs material using a 514nm laser, repetition rate 76 MHz. 

\section{Polarition condensation in single micropillar}
We demonstrate the polariton condensation in a single micropillar using energy-resolved momentum space PL. Fig.S2a and b show a single micropillar's energy-resolved momentum space PL below and above the condensation threshold. 
\begin{figure}[H]%
\centering
\includegraphics[width=0.9\textwidth]{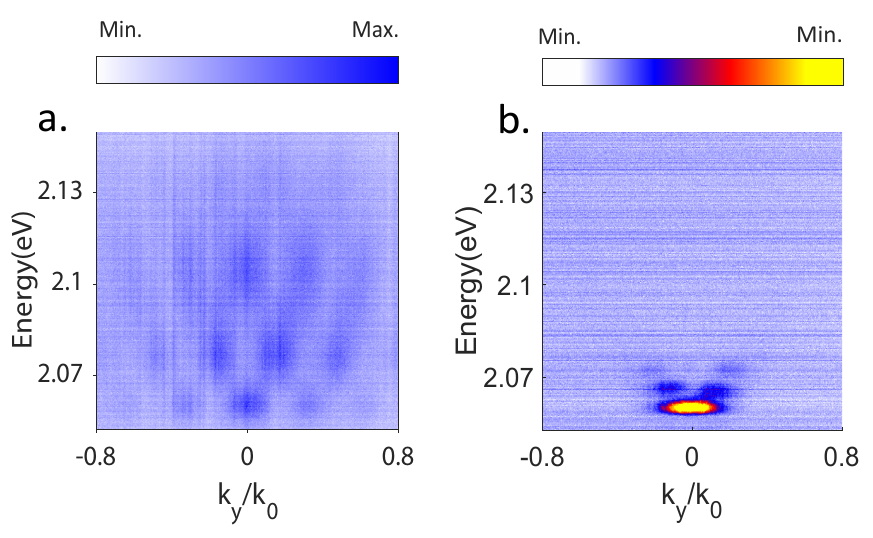}
\caption{Momentum space energy resolved photoluminescence from  micropillar with diameter,$d=2.75\mu$m  under (a) linear regime at $p<p_{th}$ and  non-linear regime at (b) $p$=$ 1.25p_{th}$\label{fig2}}
\end{figure}

\section{ Bandstructure of honeycomb polariton lattice }\label{sec1}
\begin{figure}[H]%
\centering
\includegraphics[width=0.9\textwidth]{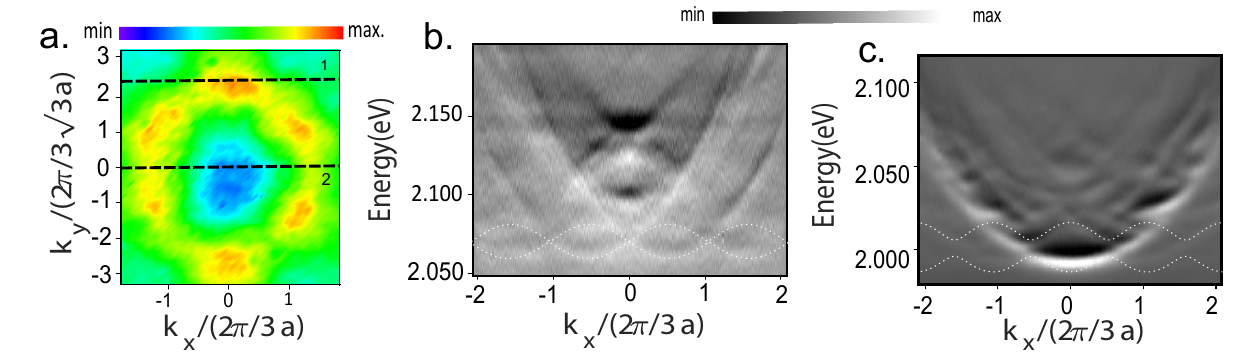}
\caption{  (a) Measured BZ of the honeycomb lattice at the Dirac energy. Experimentally measured band structure along with tight binding model fits to the observed dispersion along (b) line 1 at $k_y=4\pi/(3\sqrt{3}a)$  and (c) line 2 at $k_y=0$ in (a).}\label{figS3}
\end{figure}

 PL from the honeycomb lattice is captured using energy-resolved momentum space imaging, employing a non-resonant pulsed laser, 514nm with a repetition rate of $~76 MHz $ and a spot size of 25$\mu m$. In the low excitation limit, the incoherent relaxation of polaritons leads to the population of all energy bands. The first Brillouin zone (BZ) of the lattice is measured using FIM at Dirac energy, as depicted in Fig.\ref{figS3}a. Dirac points are observed as bright spots at the six corners of the measured first BZ. Energy bands of the honeycomb polariton lattice are measured along line 1 and line 2 (black lines in Fig.\ref{figS3}a). Fig.\ref{figS3}b displays the energy bands along line 1, exhibiting Dirac points at $k_x/k_0=0,\pm (2\pi/3a)$, accompanied by other higher-order energy bands. The measured bandstructure along line 2, beyond the Dirac point, displays energy bands with a resolved energy gap, as illustrated in Fig.\ref{figS3}c.

\end{document}